\documentclass[aps,pra,reprint, amsmath, amssymb,superscriptaddress]{revtex4-1}
 
\usepackage{bm}
\usepackage[retainorgcmds]{IEEEtrantools}
\usepackage{graphicx}
\usepackage{mathrsfs}
\usepackage{amsmath}
\usepackage{amsfonts}
\usepackage{amssymb}
\usepackage{color}
\usepackage{times,txfonts}
\usepackage{nicefrac}
\usepackage{ragged2e}
\usepackage{tikz}
\usepackage{physics}

\usepackage[colorlinks=true,linkcolor=blue,urlcolor=blue,citecolor=blue,pdfusetitle]{hyperref}

\usepackage{blindtext}

\definecolor{cadmiumgreen}{HTML}{097969}

\begin{document}

\title{Fermionic Wigner function formalism for the description of transport in chains with inhomogeneous dephasing rates}
\title{Wigner dynamics of a quantum gas under inhomogeneous losses and dephasing}
\title{Wigner dynamics for quantum gases under inhomogeneous gain and loss processes with dephasing}
\author{Michele Coppola}
\email{michele.coppola@univ-lorraine.fr}
\affiliation{Université de Lorraine, CNRS, LPCT, F-54000 Nancy, France}
\author{Gabriel T. Landi}
\email{gabriel.landi@rochester.edu}
\affiliation{Department of Physics and Astronomy, University of Rochester, Rochester, New York 14627, USA}
\affiliation{Instituto de F\'isica da Universidade de S\~ao Paulo,  05314-970 S\~ao Paulo, Brazil.}
\author{Dragi Karevski}
\email{dragi.karevski@univ-lorraine.fr}
\affiliation{Université de Lorraine, CNRS, LPCT, F-54000 Nancy, France}

\begin{abstract}
We present a Wigner function-based approach for the particle density evolution in fermionic and bosonic open quantum many-body systems, including the effects of dephasing. In particular, we focus on chains of non-interacting particles coupled to Lindblad baths. The dissipative processes, described by linear and quadratic jump operators, are modulated by inhomogeneous couplings. Following a semi-classical approach, we find the differential equation governing the Wigner function evolution, which can be solved in closed form in some particular cases. We check the accuracy of the Wigner approach in different scenarios (i.e. Gaussian jump rates), describing the density evolution and the transport phenomena in terms of classical quasi-particles. 
\end{abstract}

\maketitle{}

\section{Introduction}
Studying the particle transport in out-of-equilibrium quantum systems has always been a very attractive topic. In this direction, methods involving the Wigner function have achieved a great success in semi-classical contexts \cite{wigner1997quantum,moyal1949quantum}. The theory, based on mapping quantum observables into phase-space real-valued functions, reduces to the famous Boltzmann transport equation for the Wigner quasi-probability distribution \cite{boltzmann1872weitere,1987stme.book.....H,stoimenov2016conformal}, elegantly defined as the Weyl transform of the density operator \cite{hinarejos2012wigner,dean2018wigner,de2021wigner}. 

Despite the surge in interest, several scenarios still remain partially unexplored, such as the $k$-body gain and loss processes in one dimensional systems \cite{bouchoule2020effect,dast2014quantum,alba2022noninteracting,alba2022hydrodynamics,carollo2022dissipative}. 
Recently, a characteristic function approach has been developed to treat open fermion systems \cite{wang2022exact}, similarly to the phase-space method widely used in quantum optics \cite{carmichael1999statistical,schleich2011quantum,santos2017wigner,malouf2019wigner}. The characteristic function approach is based on building a map between the Liouville-Fock space and the Grassmann algebra. For general quadratic Hamiltonians and linear Lindbladian operators, the quantum master equation of the density matrix is transformed into a first order partial differential equation for the characteristic function, exactly solvable by standard techniques. This approach represents a valid alternative to the third quantization method \cite{prosen2008third}, thanks to the rich analytic and algebraic tools for functions in the Grassmann algebra. For instance, the average of one-body or two-body observables can be expressed by partial derivatives of the characteristic function. 

Particle transport in open Markovian systems has been attracting a lot of attention \cite{alba2021spreading,landi2021waiting,silva2022non,karevski2009quantum,landi2021waiting,prosen2011exact,karevski2013exact,popkov2013driven,landi2014flux,landi2015open}.
This work got inspiration from the hydrodynamic approach of Refs.~\cite{castro2016emergent,ruggiero2020quantum,doyon2020lecture,bastianello2019generalized,capizzi2022domain,scopa2022exact,fagotti2017higher,scopa2021exact,dubail2017conformal,collura2018analytic,collura2020domain,alba2021generalized,bouchoule2022generalized,bulchandani2017solvable,bulchandani2018bethe,doyon2018soliton,schemmer2019generalized,malvania2021generalized,collura2012entangling,wendenbaum2013hydrodynamic,cao2019entanglement,jin2021interplay}, based on the intriguing idea of describing the open dynamics in terms of classical non-interacting quasi-particles, to grasp the apparently complexity of the dissipative processes. 
In particular, our goal here is to put forth a hydrodynamic description of open process in systems combining inhomoheneous gain and loss, and dephasing. 

In particular, we consider quantum chains of free spinless fermions coupled to Lindblad baths, whose interaction is described by linear and quadratic jump operators, and can describe arbitrary emission and absorption, both in position and momentum space.
In addition, we also introduce dephasing, a type of reservoir which introduces noise in the system, but without an accompanying particle current. This type of noise is  well known in continuous measurement scenarios, in which a quantum chain of non-interacting fermions is coupled to an external monitoring apparatus detecting the local occupation \cite{cao2019entanglement,carollo2022entangled,coppola2022growth,piccitto2022entanglement,alberton2021entanglement,muller2022measurement,bernard2018transport,cai2013algebraic,tirrito2022full}. The dephasing rate coincides with the monitoring frequency or, using the generalized hydrodynamics approach, the annihilation rate of quasi-particle pairs spreading ballistically with opposite momentum. In weak measurement protocols, it has been proved that the inverse of the monitoring rate corresponds to the characteristic time at which the ballistic regime is replaced by a diffusive one \cite{cao2019entanglement,eisler2011crossover}.

In the first part of this work (Sec.~\ref{sec:gen}), we  define the Lindblad dynamics and give the equations of motion of the two-point functions. In Sec.~\ref{sec:hydro}, taking the hydrodynamic limit, we derive a Wigner
phase space representation of the dynamics of the correlation matrix which turns to be governed by a linear differential equation.  
In the semi-classical limit we show that the equation of motion admits a simple probabilistic interpretation in terms of non-interacting classical quasi-particles. We consider first the situation without
 dephasing and local potential  (Sec.~\ref{sec:no_dephasing}) for which the equation of motion is exactly solvable by standard techniques.  
We then show in Sec.~\ref{sec:dephasing} that a constant dephasing dramatically affects the Wigner function dynamics after a crossover time, with a new emerging evolution reflecting the diffusive motion of the quasi-particles through the chain. 
Our main findings are summarized in Sec.~\ref{sec:disc}, where we draw some future perspectives. 

\section{General framework}
\label{sec:gen}

Let $\hat\rho$ be the density operator of a quantum system whose dynamics is generated by the Liouvillian $\mathcal{L}(\hat{\rho})$ of the form
\begin{equation}\label{M_gen}
     \frac{d\hat{\rho}}{dt} = \mathcal{L}(\hat{\rho}) =- i [\hat{H},\hat{\rho}] + D(\hat\rho),\quad  D(\hat\rho)=\sum_j \Upsilon_j\mathcal{D}(\hat{L}_j),
\end{equation}
where $\hat H$ denotes the Hamiltonian, $D(\hat\rho)$ is the dissipator and $\mathcal{D}(\hat{L}_j)$ is the super-operator
\begin{equation}
    \mathcal{D}(\hat{L}_j)=\hat{L}_j \hat{\rho} \hat{L}_j^\dagger - \frac{1}{2} \{\hat{L}_j^\dagger \hat{L}_j, \hat{\rho}\},
\end{equation}
acting on the Linbladian operators $\hat{L}_j$. The jump operators characterize the interaction with the environment and the rates $\Upsilon_j>0$ determine the coupling strength system-bath. 

In this work, we will consider $N$-site quantum chains under periodic boundary conditions. 
Our main interest will be on spinless fermions. However, the results also hold for bosons, with minimal modifications. In what follows, we will therefore consider both statistics side by side. 
In order to express formulas in compact form, every time the commutation/anti-commutation rules induce a change of sign, the one on top refers to fermions and the one on bottom to bosons. 

We consider a quadratic Hamiltonian $\hat H=\hat H_0+\hat V$, where $\hat H_0$ is translationally invariant and $\hat V$ is assumed to be diagonal in real-space representation. In the most general case, $\hat H_0$ and $\hat V$ are not commuting observables; $\hat H_0$ describes the hopping between sites (tight-binding) and $\hat V$ plays the role of a local potential which, for instance, may depend on experimentally tunable parameters. The explicit form of the operators $\hat H_0$ and $\hat V$ is
\begin{equation}\label{H}
    \hat H_0 = \hat{\bf c}^\dagger h_0 \hat{\bf c},\quad \hat V = \hat{\bf c}^\dagger\mathcal{V} \hat{\bf c},\quad \hat{\bf c}=\begin{pmatrix}\hat{c}_1\\ \vdots \\\hat{c}_N\end{pmatrix},
\end{equation}
where the $\hat{c}$'s are the fermionic/bosonic operators and $h_0, \mathcal{V}$ are $N\times N$ Hermitian matrices. By assumption, $\mathcal{V}_{xy}=\mathcal{V}_{x}\delta_{xy}$ and, without loss of generality, we can suppose $(h_0)_{xx}=0$ $\forall x$. Therefore the full Hamiltonian is $\hat H = \hat{\bf c}^\dagger h \hat{\bf c}$, with $h=h_0+\mathcal{V}$ being the complete coefficient matrix.

By hypothesis, $\hat H_0$ is translationally invariant and may be put in diagonal form with the canonical transformation, 
\begin{equation}\label{Fourier_modes}
    \hat{\eta}_p = \frac{1}{\sqrt{N}}\sum_x e^{-ipx}\hat{c}_x,\qquad \hat{\eta}^\dagger_p = \frac{1}{\sqrt{N}}\sum_x e^{ipx}\hat{c}^\dagger_x,
\end{equation}
where the label $x$ runs over all the chain sites; the $p$'s are the so-called Fourier modes belonging to the Brillouin zone $\mathcal{B}=\{p=-\pi+2\pi l/N:l\in[0,N-1],l\in\mathbb{N} \}$; $(\hat{\eta}^\dagger,\hat{\eta})$ are the rising and lowering operators in momentum space. 

We consider the following general form for the dissipators in Eq.~\eqref{M_gen}, including both linear and quadratic jump operators:
\begin{equation}\label{dissipator}
\begin{aligned}
      D(\hat \rho) =& \sum_x \gamma^+_x\mathcal{D}(\hat{c}^\dagger_x)+ \gamma^-_x\mathcal{D}(\hat{c}_x)+
      \lambda_x\mathcal{D}(\hat{c}^\dagger_x \hat{c}_x)\\
    &+\sum_{p\in\mathcal{B}} \omega^+_p\mathcal{D}(\hat{\eta}^\dagger_p)+ \omega^-_p\mathcal{D}(\hat{\eta}_p)+\zeta_p\mathcal{D}(\hat{\eta}^\dagger_p \hat{\eta}_p).
\end{aligned}
\end{equation}
A description of each coefficient, together with its physical interpretation, is provided in Table~\ref{tab:coefficients}.
It is important to remark that all the coupling constants appearing in Eq. \eqref{dissipator} depend on the interaction Hamiltonian and the hypothesis on the bath. Ideally, one should always attempt to derive the dissipators and the jump rates starting from a microscopic theory. 
For instance, it is quite common to assume baths of bosons/fermions and linear couplings to the system. In this way, the dissipator gives physically reasonable results and reproduces some expected behaviors like the relaxation to a Gibbs thermal state. However, as expected, building a microscopic theory is not always possible and this problem is commonly avoided by using phenomenological dissipators. In the view of presenting the most general case, we will not assume any constraint on the jump rates, which will be considered independent quantities.
\begin{figure}[t!]
\begin{center}
\includegraphics[width=.37\textwidth]{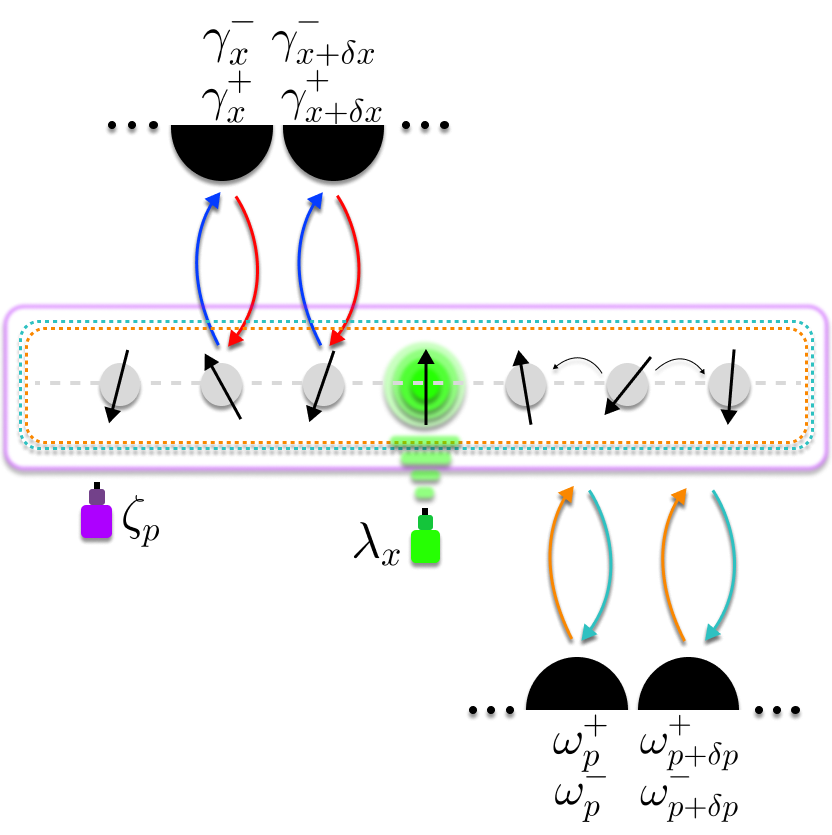}
     \caption{\label{id}\textbf{Open quantum dynamics:} Sketch of an open fermionic quantum system. We consider different dissipative effects. Rates $\gamma^+_x$, $\gamma^-_x$, $\omega^+_p$ and $\omega^-_p$ are related to local and non-local fermion gain and loss processes; $\lambda_x$ and $\zeta_p$ are the dephasing rates, corresponding to the monitoring rates of position and momentum in the formalism of quantum trajectories. }
\end{center}
\end{figure}

\begin{table}[]
    \centering
    \caption{Description of the main coefficients entering the master equation~\eqref{dissipator}.}
    \begin{tabular}{c|l}
        Coefficient & Interpretation 
        \\[0.2cm]
        \hline
        $h_0$ & Trans. inv.; Eigenvalues $\epsilon_p$
        \\[0.1cm]
        $\mathcal{V}_{xy}=\mathcal{V}_{x}\delta_{xy}$ & Diagonal potential 
        \\[0.1cm]
        $\gamma_x^+,\hspace{0.2cm}\gamma_x^-$ & Inj./ext. rate at position $x$
        \\[0.1cm]
        $\omega_p^+,\hspace{0.2cm}\omega_p^-$ & Inj./ext. rate with momentum $p$.
        \\[0.1cm]
        $\lambda_x$ & Dephasing  at position $x$
        \\[0.1cm]
        $\zeta_p$ & Dephasing  with momentum $p$
        \\[0.1cm]
        $\Gamma_x = \gamma_x^- \pm \gamma_x^+$ & Dissipation  at position $x$
        \\[0.1cm]
        $\Omega_p = \omega_p^- \pm \omega_p^+$ & Dissipation  with momentum $p$
        \\[0.3cm]
        $n(x,p,t)$ & Wigner function, Eq.~\eqref{Wigner_funct_def}
        \\[0.1cm]
        $\rho(x,t), \tilde{\rho}(p,t)$ & Densities [Eq~\eqref{position_and_momentu_densities}]
        \\[0.1cm]
        $N_p(t) = \sum_x \rho(x,t) = \sum_{p\in \mathcal{B}} \tilde{\rho}(p,t)$ 
        & Number of particles in the chain
    \end{tabular}
    \label{tab:coefficients}
\end{table}

One of the possible ways to visualize intuitively the Lindblad dynamics \eqref{M_gen} with the dissipator \eqref{dissipator} coincides with the quantum trajectory techniques \cite{daley2014quantum,wiseman1996quantum,dalibard1992wave,gardiner1992wave,carollo2019unraveling}. These techniques involve rewriting the master equation as a stochastic average over individual trajectories, which evolve in time as pure states. 
In particular, in the so-called jump unravelling, a non-Hermitian effective Hamiltonian generates a non-unitary dynamics, which is perturbed by quantum jumps randomly appearing with characteristic rates. The jump operations acting stochastically in time may involve one or more chain sites. For instance, the Lindblad operators $\hat{c}^\dagger_x$, $\hat{c}_x$ create and destroy localized particles, while $\hat{\eta}^\dagger_k$, $\hat{\eta}_k$ create and destroy fully delocalized particles, as plane waves. Finally, the quadratic Lindblad operators $\hat{c}^\dagger_x \hat{c}_x$ and $\hat{\eta}^\dagger_k \hat{\eta}_k$ represent the dephasing baths, which conserve the particle number, acting on the state like projectors. 

To obtain a hydrodynamic description, we use the Wigner function formalism to get the particle density under the Lindblad evolution \eqref{M_gen}. This allows us to cast the hydrodynamics in terms of non-interacting quasi-particles. 

Let $C_{xy}=\langle \hat{c}^\dagger_y \hat{c}_x\rangle=\tr(\hat{c}^\dagger_y \hat{c}_x\hat{\rho})$ be the elements of the so-called correlation matrix. From Eq. \eqref{M_gen} these quantities evolve according to
\begin{equation}
    \frac{d\langle \hat{c}^\dagger_y \hat{c}_x\rangle}{dt}=-i\Big\langle[\hat{c}^\dagger_y \hat{c}_x,\hat{H}]\Big\rangle + \tr\Big(\hat{c}^\dagger_y \hat{c}_x D(\hat{\rho})\Big).
\end{equation}
Thanks to the algebraic properties of the fermionic or bosonic operators the differential equation for the two-point function takes a closed from.  
To see this let us first introduce the following matrices  $\gamma^+,\gamma^-, \lambda, \tilde\omega^+, \tilde\omega^-, \tilde\zeta, \tilde\zeta^{(\alpha,\beta)}$   with elements 
\begin{subequations}
\begin{equation}
\gamma^+_{xy}=\gamma^+_x\delta_{xy},\quad\gamma^-_{xy}=\gamma^-_x\delta_{xy},\quad\lambda_{xy}=\lambda_x\delta_{xy},
\end{equation}
\begin{equation}
\tilde\omega^+_{xy}=\frac{1}{N}\sum_{p\in\mathcal{B}}e^{ip(x-y)}\omega^+_p,\quad\tilde\omega^-_{xy}=\frac{1}{N}\sum_{p\in\mathcal{B}}e^{ip(x-y)}\omega^-_p,
\end{equation}
\begin{equation}
\tilde\zeta_{xy}=\frac{1}{N}\sum_{p\in\mathcal{B}}e^{ip(x-y)}\zeta_p,\qquad\tilde\zeta^{(\alpha,\beta)}_{xy}=\tilde\zeta_{\alpha+x,\beta+y},
\end{equation}
\end{subequations}
Using the dissipator \eqref{dissipator} with some algebraic manipulations, we obtain the differential equation 
\begin{equation}\label{corr_mat_ev}
    \frac{dC_{\alpha\beta}}{dt}=- \Big(W C + C W^\dagger-F-\text{diag}(\lambda C)\Big)_{\alpha\beta}+ \frac{1}{N}\tr(\tilde\zeta^{(\alpha,\beta)}C),
\end{equation}
where 
\begin{equation}
    W = ih + (\gamma^- \pm \gamma^+ + \lambda + \tilde\omega^- \pm \tilde\omega^+ + \tilde\zeta)/2, 
\end{equation}
and $F=\gamma^+ + \tilde\omega^+$ \cite{turkeshi2021diffusion}. Eq. \eqref{corr_mat_ev} describes the exact microscopic dynamics of the correlations. As expected, in the absence of dephasing, Eq. \eqref{corr_mat_ev} reduces to the Lyapunov equation
\begin{equation}\label{lyap}
    \frac{dC}{dt}=- W C - C W^\dagger+F.
\end{equation}

In principle, the solution of Eq. \eqref{corr_mat_ev} gives everything necessary to extrapolate the time-evolved spatial density of particles. The same goes for the particle occupation in Fourier space, by taking the unitary transformation 
\begin{equation}
    \tilde C_{qp}=\langle \hat\eta^\dagger_p\hat\eta_q\rangle = \tr(\hat\eta^\dagger_p\hat\eta_q\hat{\rho})=\frac{1}{N}\sum_{xy}e^{-iqx}C_{xy}e^{ipy}.
\end{equation}
However, providing a simple physical description of the density evolution in this dissipative process is generally difficult. For this reason, our goal is to build a one-to-one map between the correlation matrix and a new real-valued function; hopefully, this will give us the insight to create a simple picture starting from the new equation of motion. The accuracy of such a picture will be tested by comparing the provided results with the exact microscopic dynamics \eqref{corr_mat_ev}.

In this view, the well-known Wigner function will play a key role in this work; it is defined as
\begin{equation}\label{Wigner_funct_def}
\begin{aligned}
 n(x,p,t)&:=\sum_{y}e^{2ipy} C_{x-y,x+y}(t)\\&=\sum_{k\in\mathcal{B}}e^{-2ixk}\tilde C_{p-k,p+k}(t),
\end{aligned}
\end{equation}
where $x$ and $p$ are the discrete position-momentum variables generating the phase-space \cite{scopa2022exact,cao2019entanglement,jin2021interplay,scopa2021exact}. The Wigner function is a joint quasi-probability distribution, and its marginals coincide with the single-particle densities in real and momentum space:
\begin{equation}\label{position_and_momentu_densities}
\begin{aligned}
    \rho_x &= \langle\hat c^\dagger_x \hat c_x\rangle = \frac{1}{N}\sum_{p\in\mathcal{B}} n(x,p,t),
    \\[0.2cm]
    \tilde\rho_p &= \langle\hat \eta^\dagger_p \hat \eta_p\rangle=  \frac{1}{N}\sum_{x} n(x,p,t),
\end{aligned}
\end{equation}
which follow from Kronecker delta relations $\delta_{xy}=(1/N)\sum_{k\in\mathcal{B}}e^{ik(x-y)}$ and $\delta_{pq}=(1/N)\sum_{x}e^{ix(p-q)}$.
The total number of particles spreading on the chain is $N_p(t)=(1/N)\sum_{x}\sum_{p\in\mathcal{B}} n(x,p,t)$, which is an extensive quantity. 

In the absence of dephasing, the Lindblad dynamics is Gaussian preserving. In such a case, if the system is properly prepared in a Gaussian state at time $t=0$ then the Wigner function \eqref{Wigner_funct_def} provides a complete description of the quantum dynamics. Indeed, Gaussian states are fully characterized by the two point functions and the Wick theorem provides any many particle quantity.

\section{Hydrodynamics}
\label{sec:hydro}

The next goal is getting the equation of motion of the Wigner function from Eq. \eqref{corr_mat_ev}. We will proceed by evaluating each single contribution coming from~\eqref{corr_mat_ev} for the correlation matrix elements. According to Eqs. \eqref{M_gen} and \eqref{Wigner_funct_def},  
\begin{equation}\label{RHS_imp}
    \partial_t n(x,p,t)=\sum_ye^{2ipy}\tr\Big\{\hat{c}^\dagger_{x+y} \hat{c}_{x-y}\mathcal{L}(\hat{\rho}(t))\Big\}.
\end{equation}
Let's start from the unitary contribution describing the dynamics of closed systems. By hypothesis, the Hamiltonian $\hat H_0$ is diagonal in the Fourier space, 
\begin{equation}\label{h_xy}
    (h_0)_{xy}=\bra{0}\hat c_x\hat H_0 \hat c^\dagger_y\ket{0}=\frac{1}{N}\sum_{p\in\mathcal{B}}e^{ip(x-y)}\epsilon_p,
\end{equation}
where $\ket{0}$ is the vacuum state and $\epsilon_p$ are the single particle eigenvalues of $\hat H_0$. 
\begin{widetext}
Combining relations \eqref{Fourier_modes}, \eqref{h_xy} with \eqref{Wigner_funct_def}, 
\begin{multline}\label{cont1}
    -i\sum_{y}e^{2ipy}\tr\Big\{[\hat{c}^\dagger_{x+y} \hat{c}_{x-y},\hat{H}]\hat{\rho}(t)\Big\}=
    -i\sum_{k\in\mathcal{B}}e^{-2ikx}(\epsilon_{p-k}-\epsilon_{p+k})\tilde C_{p-k,p+k}(t)
    -i\sum_{y}e^{2ipy}(\mathcal{V}_{x-y}-\mathcal{V}_{x+y})C_{x-y,x+y}(t).
\end{multline}
Proceeding similarly for the dissipative part  containing linear jump operators,
\begin{multline}\label{cont2}
    \sum_{y}e^{2ipy}\sum_z\tr\bigg\{\hat{c}^\dagger_{x+y} \hat{c}_{x-y}\Big(\gamma_z^+\mathcal{D}[\hat{c}^\dagger_z]+\gamma_z^-\mathcal{D}[\hat{c}_z]\Big)\bigg\}=\gamma^+_x-
    \frac{1}{2}\sum_{y}e^{2ipy}\bigg[\gamma^-_{x-y}+\gamma^-_{x+y}\pm(\gamma^+_{x-y}+\gamma^+_{x+y})\bigg] C_{x-y,x+y}(t),
\end{multline}
and
\begin{multline}\label{cont3}
    \sum_{y}e^{2ipy}\sum_{k\in\mathcal{B}}\tr\bigg\{\hat{c}^\dagger_{x+y} \hat{c}_{x-y}\Big(\omega_k^+\mathcal{D}[\hat{\eta}^\dagger_k]+\omega_k^-\mathcal{D}[\hat{\eta}_k]\Big)\bigg\}=\omega^+_p-
    \frac{1}{2}\sum_{k\in\mathcal{B}}e^{-2ikx}\bigg[\omega^-_{p-k}+\omega^-_{p+k}\pm(\omega^+_{p-k}+\omega^+_{p+k})\bigg]\tilde C_{p-k,p+k}(t).
\end{multline}
Concerning the dephasing, one obtains 
\begin{align}\label{cont4}
    \sum_{y}e^{2ipy}\sum_{z}\tr\bigg\{\hat{c}^\dagger_{x+y} \hat{c}_{x-y}\Big(\lambda_z\mathcal{D}[\hat{c}^\dagger_z\hat{c}_z]\Big)\bigg\}=
    \lambda_x C_{xx}(t)-\frac{1}{2}\sum_{y}e^{2ipy}(\lambda_{x-y}+\lambda_{x+y}) C_{x-y,x+y}(t),
\end{align}
and
\begin{align}\label{cont5}
    \sum_{y}e^{2ipy}\sum_{k\in\mathcal{B}}\tr\bigg\{\hat{c}^\dagger_{x+y} \hat{c}_{x-y}\Big(\zeta_k\mathcal{D}[\hat{\eta}^\dagger_k\hat{\eta}_k]\Big)\bigg\}=
    \zeta_p\tilde C_{pp}(t)-\frac{1}{2}\sum_{k\in\mathcal{B}}e^{-2ikx}(\zeta_{p-k}+\zeta_{p+k})\tilde C_{p-k,p+k}(t).
\end{align}

Up to now, any calculation has been exactly solved without any approximation. In the following, we will go to a continuous limit for the position-momentum variables. Since hydrodynamics applies for mesoscopic scales, we assume the Wigner function as well as the jump rates, $\epsilon_p$ and $\mathcal{V}_x$ to be slowly varying functions in the microscopic scale, characterized by the lattice spacing and the distance $dp=2\pi/N$ between two consecutive Fourier modes. In the thermodynamic limit ($N\to\infty$), position and momentum become de facto continuous variables and the Wigner function $n(x,p,t)$ and all the quantities $\epsilon_p,\mathcal{V}_x,\gamma^+_x,\gamma^-_x,\lambda_x,\omega^+_p,\omega^-_p, \zeta_p$ smooth analytical functions with domain in the phase space $(x,p)$. 


For analytic functions, we can expand in powers of $k$ and $y$
\begin{subequations}
\begin{equation}\label{sviluppo1}
    \frac{g_{p-k}+(-1)^jg_{p+k}}{2}=(-1)^j\sum_{n=0}^{\infty}\frac{\partial^{2n+\delta_{j,1}}g_p}{(2n+\delta_{j,1})!}k^{2n+\delta_{j,1}}, 
\end{equation}
\begin{equation}\label{sviluppo2}
    \frac{f_{x-y}+(-1)^jf_{x+y}}{2}=(-1)^j\sum_{n=0}^{\infty}\frac{\partial^{2n+\delta_{j,1}}f_x}{(2n+\delta_{j,1})!}y^{2n+\delta_{j,1}}, 
\end{equation}
\end{subequations}
for $j\in\{0,1\}$, $f_x=\mathcal{V}_x,\gamma^+_x,\gamma^-_x, \lambda_x$ and $g_p=\epsilon_p, \omega^+_p,\omega^-_p,\zeta_p$. 
In this limit, the real and momentum space densities are 
 \begin{equation}
       \rho(x,t) = \frac{1}{2\pi}\int_{-\pi}^{\pi} dp\hspace{0.05cm} n(x,p,t),\qquad
     \tilde\rho(p,t) = \frac{1}{N}\int dx\hspace{0.05cm} n(x,p,t),   
 \end{equation}
with total number of particles
\begin{equation}
    N_p=\int dx \hspace{0.1cm}\rho(x)=\frac{N}{2\pi}\int_{-\pi}^{\pi} dp\hspace{0.1cm}\tilde\rho(p)=\frac{1}{2\pi}\int dx\int_{-\pi}^{\pi}dp \hspace{0.1cm}n(x,p,t).
\end{equation}
Finally, using Eqs. \eqref{sviluppo1}, \eqref{sviluppo2} in \eqref{cont1}-\eqref{cont5}, one obtains 
\begin{align}\label{Hydro_eq}
    \partial_t n(x,p,t)&= 2\Big(\epsilon_p+\mathcal{V}_x\Big)\sin(\frac{1}{2}\Big(\overset{\leftarrow}{\partial}_x\overset{\rightarrow}{\partial}_p-\overset{\leftarrow}{\partial}_p\overset{\rightarrow}{\partial}_x\Big))n(x,p,t)
    -(\omega^-_p\pm\omega^+_p+\zeta_p+\gamma^-_x\pm\gamma^+_x+\lambda_x)\cos(\frac{1}{2}\Big(\overset{\leftarrow}{\partial}_x\overset{\rightarrow}{\partial}_p-\overset{\leftarrow}{\partial}_p\overset{\rightarrow}{\partial}_x\Big))n(x,p,t)
    \nonumber
    \\[0.2cm]
    &+\int dx'\int_{-\pi}^{\pi}dp'\bigg(\frac{\zeta_{p'}}{N}\delta(p-p')+\frac{\lambda_{x'}}{2\pi}\delta(x-x')\bigg)n(x',p',t)+\gamma^+_x+\omega^+_p,
\end{align}
which is a linear differential equation in $n(x,p,t)$.
The arrows indicate the direction of differentiation and $\sin(\frac{1}{2}\Big(\overset{\leftarrow}{\partial}_x\overset{\rightarrow}{\partial}_p-\overset{\leftarrow}{\partial}_p\overset{\rightarrow}{\partial}_x\Big))$ is the well-known Moyal product \cite{wigner1997quantum,moyal1949quantum,fagotti2020locally,fagotti2017higher}. 
Eq. \eqref{Hydro_eq} describes the time evolution of the Wigner function. By hypothesis, all the terms involved in Eq. \eqref{Hydro_eq} are slowly varying functions of position and momentum. By neglecting higher order derivatives, one gets
\begin{equation}\label{Hydro_eq_21}
    \partial_t n(x,p,t)=\{\epsilon_p+\mathcal{V}_x,n(x,p,t)\}_{\mathcal{PB}}+\gamma^+_x+\omega^+_p
    -\Big(\Omega_p+\zeta_p+\Gamma_x+\lambda_x\Big)n(x,p,t)
    +\zeta_p\tilde\rho(p,t)+\lambda_x\rho(x,t),
\end{equation}
where $\Omega_p=\omega^-_p\pm\omega^+_p$, $\Gamma_x=\gamma^-_x\pm\gamma^+_x$ and 
\begin{equation}
    \{\mathcal{F},\mathcal{G}\}_{\mathcal{PB}}=\partial_x\mathcal{F}\partial_p\mathcal{G}-\partial_p\mathcal{F}\partial_x\mathcal{G}, 
    \qquad 
    \forall\mathcal{F},\mathcal{G}
\end{equation} 
indicates the Poisson bracket, describing the closed dynamics at the lower order in the $\partial_x$, $\partial_p$ derivatives \cite{fagotti2017higher,fagotti2020locally,moyal1949quantum,de2021wigner,dean2019nonequilibrium}. For a recap of the meaning of each parameter, see Table~\ref{tab:coefficients}. 

Eq.~\eqref{Hydro_eq_21} is our main result. It provides a compact partial differential equation for the Wigner function, describing inhomogeneous gain and loss terms, in both position and momentum space, as well as dephasing. Observe the presence of non local terms in Eq.~\eqref{Hydro_eq_21}, making it difficult to find the full analytical solution. Unlike the case for isolated systems, the external environment introduces terms of zero order in the $\partial_x$, $\partial_p$ derivatives.

Regarding the range of applicability of the truncation of higher order derivatives, Eq.~\eqref{Hydro_eq} suggests that the approximation is more inefficient for open systems. Indeed, in absence of external baths, the Wigner dynamics may be studied by the classical evolution plus quantum corrections at the third order in the partial derivatives. On the other hand, Eq.~\eqref{Hydro_eq_21} neglects contributions of the second order in the partial derivatives, involving jump rates. As a consequence of this, the reasonableness of the truncation operation needs to be evaluated more carefully, case by case.
\end{widetext}

Eq. \eqref{Hydro_eq_21} can be described as a stochastic average over individual trajectories of classical particles. In the same spirit of the generalized hydrodynamic description (GHD), weakly-entangled but highly excited initial states behave like a reservoir of classical non-interacting quasi-particles. For closed systems, at $t>0$ the particle dynamics is governed by Newton’s laws and the total number of particles is a constant of motion ($[\hat{H},\hat{N}_p]=0$). For instance, in the absence of local potentials, the particles spread ballistically with group velocity $v_p=\partial_p \epsilon_p$. A non zero local potential breaks the momentum conservation and any excitation in the phase-space point $(x_0,p_0)$ at time $t_0$ moves to $(x_0+v_{p_0}dt,p_0+\mathcal{F}_{x_0}dt)$ at time $t_0+dt$, with $\mathcal{F}_x=-\partial_x\mathcal{V}_x$. 

In the interest of clarity, we will discuss the effects of the dissipators one by one, starting from the single particle gain and loss processes and concluding with the dephasing. The linear jump operators affect the Winger evolution in two different ways. The first important difference compared to the closed dynamics regards the particle life-time. According to the hydrodynamic equation \eqref{Hydro_eq_21}, the classical particles crossing the phase-space point $(x,p)$ are destroyed with local frequency $\alpha_{(x,p)}=\Gamma_x+\Omega_p$. The other consequence derives from the source term $\gamma^+_x+\omega^+_p$, being the creation rate of new excitations in the phase space: $\gamma^+_x dt$ represents the probability in every interval $dt$ of injecting quasi-particles at position $x$ and random momentum $p$ uniformly distributed in the Brillouin zone; $\omega^+_p dt$ is the probability of creating quasi-particles with momentum $p$ and random site $x$ uniformly distributed along the chain. One fundamental aspect of the open dynamics emerges: the presence of only linear jump operators cannot change the transport features. The motion of each classical particle is still governed by Newton’s laws and the dissipators only add or remove excitations in the phase space. The possible addition of dephasing terms changes that, and makes the dynamics much more interesting. The particle life-time further reduces, with destroying frequency $\alpha'_{(x,p)}=\alpha_{(x,p)}+\lambda_x+\zeta_p$. However, there are also the last two non local terms $\lambda_x\rho(x,t)$ and $\zeta_p\tilde\rho(p,t)$ in Eq.~\eqref{Hydro_eq_21}.  For each quasi-particle crossing the site $x$, a new excitation is created with frequency $\lambda_x$, at position $x$ and random momentum $p$ uniformly distributed in the Brillouin zone. Finally, for each quasi-particle with momentum  $p$, a new excitation is created with frequency $\zeta_p$, momentum $p$ and random site $x$ uniformly distributed in $(0,N)$. Quadratic jump operators strongly affect the transport features: the pure and homogenous dephasing case ($\lambda_x=\lambda$) is explicative from this point of view, where the single-particle density $\rho(x,t)$ satisfies a Fokker-Planck differential equation 
\begin{equation}\label{diffusione}
    \partial_t\rho(x,t)=\frac{D}{2}\partial_{xx}\,\rho(x,t),
\end{equation}
with diffusive coefficient $D=\lambda^{-1}$, in the large $\lambda$ limit \cite{cao2019entanglement,eisler2011crossover}. 

The quasi-particle approach for open systems represents the second main results of this work. In the next sections, we will test the picture in some concrete examples, where the quasi-particle motion clearly emerges in the phase-space.

\section{Wigner dynamics without dephasing}
\label{sec:no_dephasing}

In order to gain some intuition on the physics of the problem, we first focus on the Wigner function evolution \eqref{Hydro_eq_21} without local potentials ($\mathcal{V}_x=0$) and dephasing ($\zeta_p=\lambda_x=0$). This choice is motivated by the huge interest in the transport regimes of boundary-driven spin chains, where the fermionic jump operators determine the exchange of excitations with the external environment. On the other hand, in quantum optics, inhomogenous single-particle gain-loss processes may be engenereed by experimentalists. For instance, coherent bosonic dynamics may be perturbed by electonic beams to remove atoms from selected sites \cite{wurtz2009experimental}, while local excitations may be created by the Raman pumping process \cite{schneble2004raman}.

Under these assumptions, Eq. \eqref{Hydro_eq_21} reduces to
\begin{equation}\label{Hydro_eq_nodep}
    \partial_t n(x,p,t)=-v_p\partial_xn(x,p,t)+\gamma^+_x+\omega^+_p
    -\Big(\Omega_p+\Gamma_x\Big)n(x,p,t),
\end{equation}
with analytical solution 
\begin{multline}\label{solution_cauchy}
 n(x,p,t)=\exp{-\Omega_p t-\int_0^{t} dt_1\hspace{0.1cm}\Gamma(x-t_1v_p)}n(x-tv_p,p,0)\\
   +\int_0^t dt_1\gamma^+(x-t_1v_p)
   \exp{-\Omega_pt_1-\int_{0}^{t_1}dt_2\Gamma(x-t_2v_p)}\\
   +\omega^+_p\int_0^t dt_1\exp{-\Omega_pt_1-\int_{0}^{t_1}dt_2\Gamma(x-t_2v_p)},
\end{multline}
where $n(x,p,0)$ is the local density at time $t=0$, which is assumed to be known by hypothesis. We now illustrate this within specific examples. 

\subsection{Jump operators creating/destroying delocalized particles}
Let us suppose $\gamma^+_x=\gamma^-_x=0$, so that particles can be injected/ejected only as plane waves (with rates $\omega_p^+$ and $\omega_p^-$). 
The evolution of the Wigner function is given by 
\begin{equation}\label{Hydro_eq_3}
    \partial_t n(x,p,t)=-v_p\partial_x n(x,p,t)-\Omega_p n(x,p,t)+\omega^+_p,
\end{equation}
with explicit solution
\begin{equation}\label{Wigner_coupling_momentum}
    n(x,p,t)=
    \begin{cases}
    e^{-\Omega_p t}\Big(n(x-tv_p,p,0)-\frac{\omega^+_p}{\Omega_p}\Big)+\frac{\omega^+_p}{\Omega_p}\\[0.2cm]
    n(x-tv_p,p,0)+\omega^+_p t\hspace{1.0cm} \Omega_p= 0
    \end{cases}.
\end{equation}
For bosons, $\Omega_p = \omega_p^- - \omega_p^+$ and, in order for the system to be stable, we must have $\omega_p^- > \omega_p^+$.
For fermions $\Omega_p=\omega^+_p+\omega^-_p\geq 0$ and no such restriction applies.   
According to Eq. \eqref{Wigner_coupling_momentum}, the system exponentially reaches a steady state for $\Omega_p> 0$. In the long time limit, the filling factor is
\begin{equation}\label{Wigner_stead_momentum}
    \kappa_\infty:=\lim_{t\to\infty}\frac{N_p(t)}{N}=\frac{1}{2\pi}\int_{-\pi}^{\pi}dp\frac{\omega^+_p}{\Omega_p}.
\end{equation}
If there are no emissions in a fermion system, $\omega^+_p\neq 0$, $\omega^-_p=0$ and the filling factor goes to $1$. This is an example of trivial evolution where the system starts absorbing particles until all sites are full and the Pauli principle freezes the dynamics. On the other hand, if $\omega^+_p= 0$ and $\omega^-_p\neq0$, the Wigner function goes exponentially to zero and the system reaches the vacuum steady state. Another interesting case is $\omega^+_p=\omega^-_p\neq 0$, where the system tends to a maximally mixed state with particle number $N_p=N/2$, which is the condition of half filling.

Finally, if $\omega_p^+=\omega^+,\omega_p^-=\omega^-$ are $p$-independent then $\kappa_\infty=\chi/(1\pm\chi)$, where $\chi=\omega^+/\omega^-$ is the ratio between the absorption and emission rates. For bosonic systems, the asymptotic filling factor is defined only for $\chi\in[0,1)$, which is the condition for the dynamical equilibrium.

\subsection{Jump operators creating/destroying localized particles}

Next we discuss local gain and losses in real space ($\omega^+_p=\omega^-_p=0$). The differential equation \eqref{Hydro_eq_nodep} reduces to 
\begin{equation}\label{Hydro_eq_3}
    \partial_t n(x,p,t)=-v_p\partial_x n(x,p,t)-\Gamma_x n(x,p,t)+\gamma^+_x,
\end{equation}
with analytical solution
\begin{multline}\label{solution_cauchy_position}
 n(x,p,t)=\exp{-\int_0^{t} dt_1\hspace{0.1cm}\Gamma(x-t_1v_p)}n(x-tv_p,p,0)\\
   +\int_0^t dt_1\gamma^+(x-t_1v_p)
   \exp{-\int_{0}^{t_1}dt_2\Gamma(x-t_2v_p)}.
\end{multline}
Observe that, for all the examples collected here, we will take $H_0$ to be a tight-binding Hamiltonian with nearest-neighbors only; 
viz., $(h_0)_{ij}=-(\delta_{i,j+1}+\delta_{i,j-1})/2$, which leads to single particle eigenvalues $\epsilon_p=-\cos(p)$ and group velocity $v_p=\partial_p \epsilon_p = \sin(p)$. 
\begin{figure}[t]
\begin{center}
\includegraphics[width=\columnwidth]{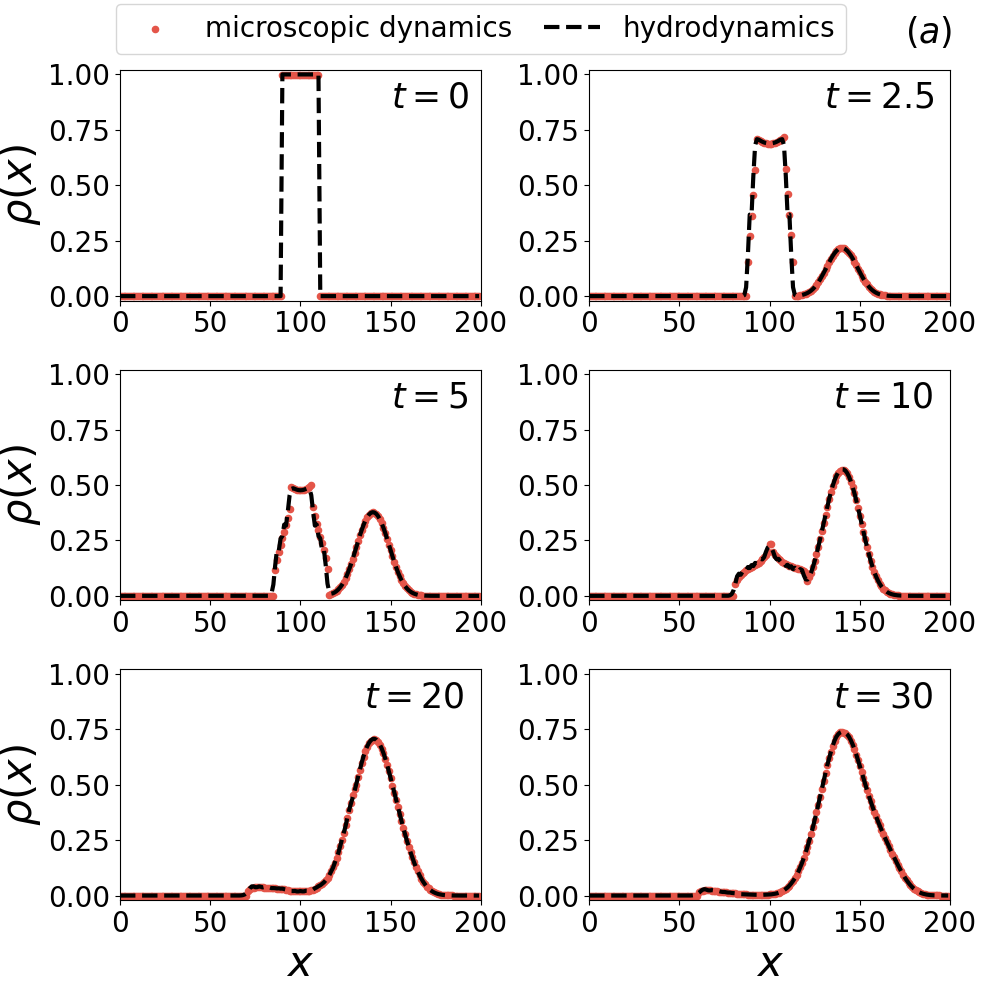}\\
\hspace{1cm}
\includegraphics[width=\columnwidth]{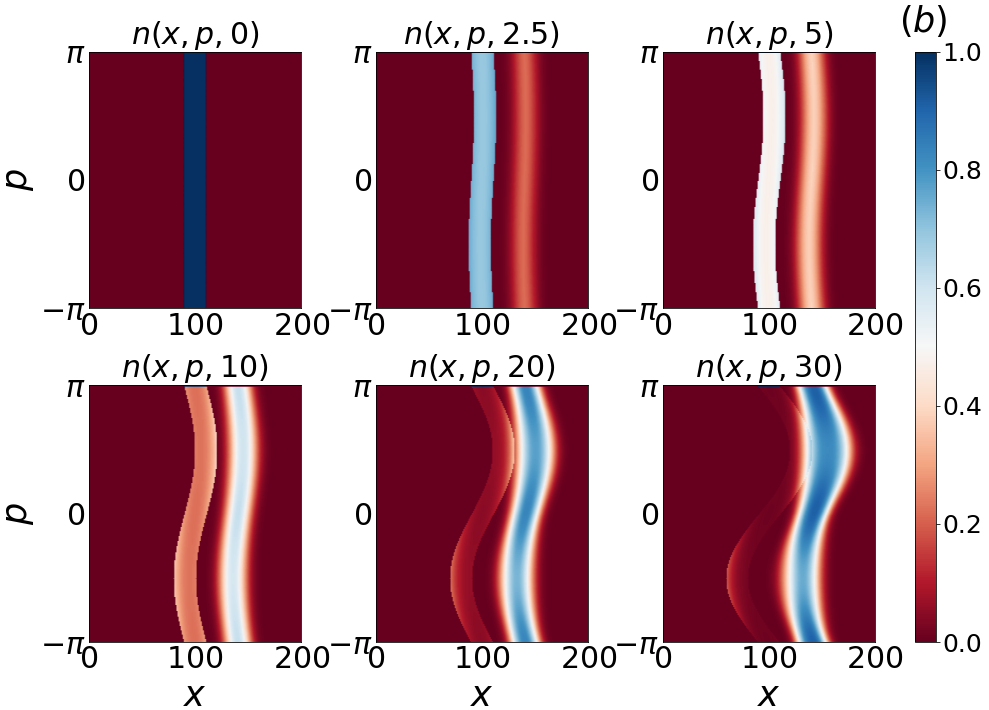}
    \caption{\label{Fermions_gauss_plots}\textbf{Density time evolution - gain and loss processes:} 200-site chain of hopping fermions with Gaussian-like jump rates \eqref{gaussian_jump_freq}. Here we set $A_+=2.0, \sigma_{x,+}=8.0, x_+=140.0, A_-=6.0, \sigma_{x,-}=16.0, x_-=100.0$. We prepare the system in the Wigner function \eqref{Wigner_step} with $\delta=10$. $(a)$ Time evolution of the density $\rho(x,t)$; the red spots and the black dashed line refer to the microscopic dynamics \eqref{lyap} and the hydrodynamic approach \eqref{solution_cauchy_position}, respectively. As expected, this particular choice of parameters generates a net flux of fermions in the RHS of the chain. In the inset, we compare the Lyapunov solution with the classical quasi-particle ansatz (blue line), by generating stochastic trajectories according to the protocol of Sec.~\ref{sec:hydro}. $(b)$ Full evolution of the phase-space Wigner function. }
\end{center}
\end{figure}

In Fig. \ref{Fermions_gauss_plots}, we consider the dynamics for hopping fermions and emission/absorption processes modulated by 
\begin{equation}\label{gaussian_jump_freq}
    \gamma^\alpha_x=\frac{A_\alpha}{\sqrt{2\pi}\sigma_{x,\alpha}}\exp{-\frac{(x-x_\alpha)^2}{2\sigma_{x,\alpha}^2}},\qquad\alpha=+,-
\end{equation}
where $A_\alpha, \sigma_{x,\alpha}, x_\alpha$ play the role of three experimental parameters. In particular, $A_\alpha$ is related to the coupling amplitude, the standard deviation $\sigma_{x,\alpha}$ introduces a dispersion along the real axis and finally $x_\alpha$ defines the peak position. We prepare the initial state
\begin{equation}\label{Wigner_step}
    n(x,p,0)=\Theta(N/2+\delta-x)-\Theta(N/2-\delta-x).
\end{equation}
For the particular choice of parameters in Figure \ref{Fermions_gauss_plots}, one can intuitively expect a net flux of particles in the RHS of the chain. In Fig. \ref{Fermions_gauss_plots} (a), we plot the density $\rho(x,t)$ as function of the position $x$ at time $t$, comparing the analytical result \eqref{solution_cauchy_position} with the solution of the Lyapunov equation \eqref{lyap}.

Once the dynamics starts,  the dissipator will begin to eject particles from the double domain wall, which is melting inside the light cones $|x-N/2-\delta|<t$ and $|x-N/2+\delta|<t$.  The particle life time $t_{\text{life}}$ is a random variable following the exponential distribution $P(t_{\text{life}};x_0,p_0)=\mathcal{Z}\exp{-\int_0^{t_{\text{life}}}ds\hspace{0.1cm} \Gamma(x_0+v(p_0)s)}$, where $\mathcal{Z}$ is a normalization constant and $x_0$, $p_0$ are the initial quasi-particle position and momentum. In this scenario, the average life time $\Bar{t}_{\text{life}}(x_0,p_0)=\int_0^\infty t P(t;x_0,p_0)dt$ is not the same for any excitation. At the same time, the source term $\gamma^+_x$ is perturbing the dynamics, with the net result of creating new excitations on the RHS of the chain. Let $t_+$ be the waiting time between two consecutive events of particle creation at position $x$: $t_+$ is a stochastic variable following the exponential distribution $Q(t_+;x)=\gamma^+_x e^{-t_+\gamma^+_x}$. In average, a new quasi-particle is created at position $x$ after any time interval $(\gamma^+_x)^{-1}$. As a consequence of the particular jump frequency profiles, the particle density becomes more and more asymmetric over time. In fact, only a few residual particles with negative velocity are able to cross the chain section dominated by the dissipation rate $\Gamma_x$ and spread ballistically to the LHS. 
In Fig. \ref{Fermions_gauss_plots} (b), we can appreciate the full Wigner function evolution, both determined by the dispersion law $\epsilon_p=-\cos(p)$ and the jump rates \eqref{gaussian_jump_freq}. 
\begin{figure}[t]
     \includegraphics[width=\columnwidth]{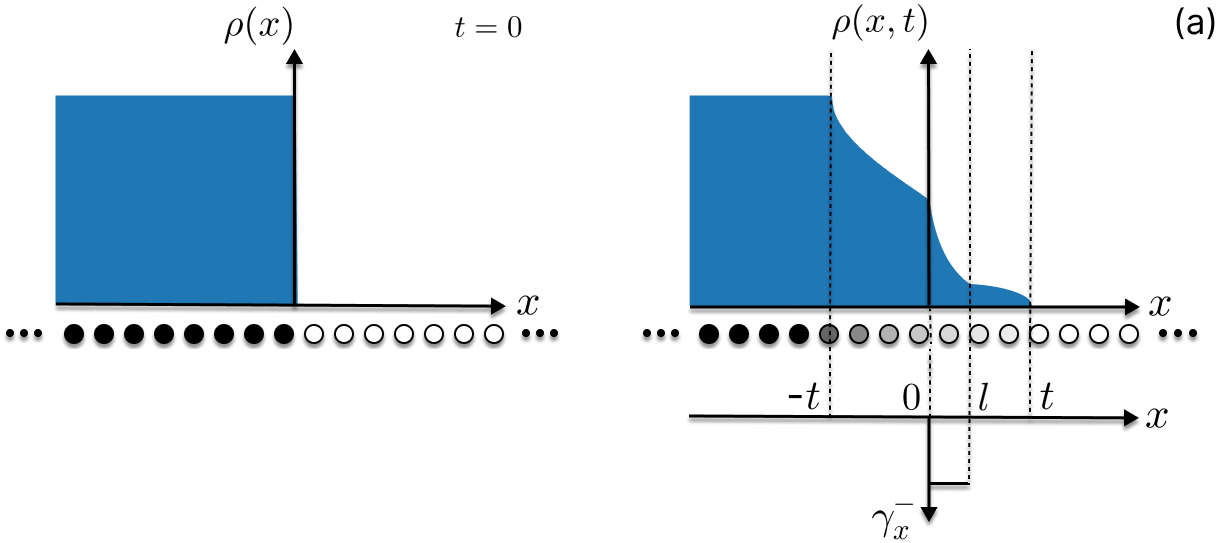}\\
     \hspace{1cm}
     \includegraphics[width=\columnwidth]{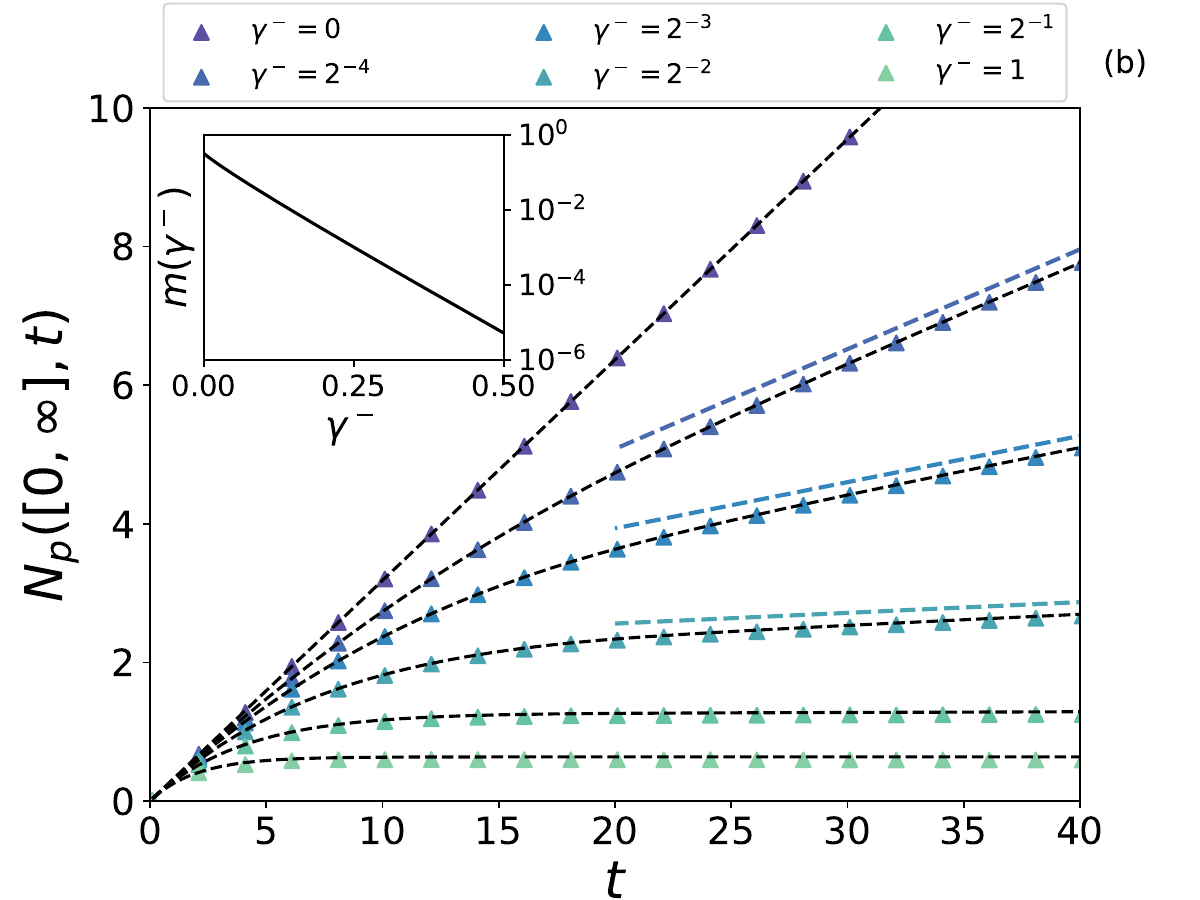}
     \caption{\label{transport2}\textbf{Density evolution - piecewise function $\gamma^-_x$:} (a) Illustration of the domain wall setting. At $t=0$ the system is entirely filled on the LHS and empty on RHS; the fermionic density is $\rho(x) = \Theta(-x)$ and the dissipative process is modulated by the jump frequency \eqref{gamma-}. At $t > 0$ the domain wall melts inside the light cone region $\abs{x}\leq t$ and the system develops a non-homogeneous density profile given by Eq.  \eqref{equazione_tran}. (b) Number of particles in $[0,\infty]$ for $x_0=l/2=10$. The dashed black line and the colored spots refer to the microscopic dynamics \eqref{lyap} and the quasi-particle ansatz \eqref{numb_0_inf}, respectively. In the inset, the coefficient \eqref{prop_coeff} for $l=20$.}  
\end{figure}

In Fig. \ref{transport2} (a), we pictorially show the density evolution for a domain wall initial configuration $n(x,p,t)=\Theta(-x)$. This choice was motivated by the large interest in out-of-equilibrium physics, where such a setup may be prepared with the help of confining potentials. After quenching the state, the transport and the spread of correlations have been studied under unitary dynamics \cite{scopa2022exact,collura2020domain,wendenbaum2013hydrodynamic}. In Fig. \ref{transport2} (a), we assume an open dynamics with single-particle loss process ($\gamma^+_x=0$). Eq. \eqref{Hydro_eq_3} reduces to
\begin{equation}\label{equazione_tran}
    \partial_t n(x,p,t)=-v_p\partial_x n(x,p,t)-\gamma^-_x n(x,p,t).
\end{equation}
We consider the annihilation rate
\begin{equation}\label{gamma-}
    \gamma^-_x=\gamma^-\bigg(\Theta(x_0+l/2-x)-\Theta(x_0-l/2-x)\bigg)\,,
\end{equation}
where $l$ is the length of the subsystem with non zero rate $\gamma^-_x$ and $x_0=l/2$. After the initial preparation, the particles start flowing towards the empty sites in the RHS of the chain. The physical picture is very simple: any particle in $[0,l]$ is destroyed with rate $\gamma^-$. The larger the annihilation rate $\gamma^-$, the higher the probability to destroy classical excitations in the phase-space.
We can use Eq.~\eqref{solution_cauchy_position} to compute $N_p([0,\infty],t)$, the number of particles in $[0,\infty]$ at time $t$ or equivalently, the total number of excitations in $[0,\infty]$ with positive group velocity and survived up to time $t$. After some algebraic manipulations, we find
\begin{multline}\label{numb_0_inf}
    N_p([0,\infty],t)=\Theta(l-t)\frac{1}{\pi\gamma^-}(1-e^{-\gamma^-t})+\Theta(t-l)\frac{1}{\pi\gamma^-}\bigg[1\\
    -e^{-\gamma^-t}\bigg(1-\sqrt{1-\bigg(\frac{l}{t}\bigg)^2}\bigg)-\int_{\arcsin(l/t)}^{\pi/2}dp\hspace{0.1cm}(\sin(p)+l\gamma^-)e^{-\frac{\gamma^-l}{\sin(p)}}\\
    +\gamma^-t\int_{\arcsin(l/t)}^{\pi/2}dp\hspace{0.1cm}\sin(p)e^{-\frac{\gamma^-l}{\sin(p)}}\bigg].
\end{multline}
This is shown in Fig. \ref{transport2} (b). 
For large times ($t/l\gg 1$), the number of particles $N_p([0,\infty],t)$ grows linearly in time, with proportionality constant 
\begin{equation}\label{prop_coeff}
    m(\gamma^-l)=\frac{1}{\pi}\int_0^{\pi/2}dp\hspace{0.1cm}\sin(p)e^{-\frac{\gamma^-l}{\sin(p)}}.
\end{equation}
If $\gamma^-l\ll 1$ then $\exp{-\frac{\gamma^-l}{\sin(p)}}\simeq 1$ for $p\in(0,\pi/2)$. In such a case, $m\simeq 1/\pi$. In the opposite limit $\gamma^-l\gg 1$, $\exp{-\frac{\gamma^-l}{\sin(p)}}\simeq 0$ and $m\simeq 0$. The inset of Fig. \ref{transport2} (b) shows the coefficient of the linear growth for $l=20$, going quickly to zero for large $\gamma^-$. Observe that, for unitary evolution, $m(0)=1/\pi$.

\subsection{Transport phenomena}

To connect the previous analysis with transport features, we will consider one single localized particle with Wigner function $n(x,p,0)=\delta(x)$. After quenching the state, we evaluate the average displacement $d(t)=\sqrt{\langle x^2\rangle - \langle x\rangle^2}$, which measures the deviation of the particle position with respect to the origin over time. As is well known, for closed systems the ballistic regime dominates the dynamics and $d(t)\propto t$. Here we would like to explore the transport for open systems, where the linear jump operators affect the particle number conservation. In the following, the dynamics verifies
Eq. \eqref{equazione_tran} with $v_p=\sin(p)$ and $\gamma^+_x=0$ to avoid particle injection. We also consider symmetric jump rates $\gamma^-_x$ with respect to $x=0$, so that $\langle x\rangle=0$ and 
\begin{equation}
    d(t)=\bigg[\int_{-\infty}^{+\infty}dx\hspace{0.1cm}x^2\rho(x,t)\bigg]^{1/2}.
\end{equation}
Solving the dynamics, 
\begin{equation}
    \rho(x,t)=
    \frac{1}{\pi t}\frac{e^{-t\int_0^1 ds\hspace{0.1cm}\gamma^-(x-xs)}}{\sqrt{1-(x/t)^2}}\bigg(\Theta(t-x)-\Theta(-t-x)\bigg).
\end{equation}
Suppose $\gamma^-_x$ is given by Eq.~\eqref{gamma-}, with $x_0=0$. In such a case,
\begin{multline}
    d(t)=\frac{t}{\sqrt{2}}e^{-\gamma^- t/2}\Theta\bigg(\frac{l}{2}-t\bigg)+\sqrt{\frac{2}{\pi}}t\bigg[\int_{\frac{l}{2t}}^{1}dy\hspace{0.1cm}\frac{y^2}{\sqrt{1-y^2}}e^{-\gamma^- l/2y}\\
    +\frac{1}{2}e^{-\gamma^- t}\bigg(\arcsin(\frac{l}{2t})-\frac{l}{2t}\sqrt{1-\bigg(\frac{l}{2t}\bigg)^2}\bigg)\bigg]^{1/2}\Theta\bigg(t-\frac{l}{2}\bigg).
\end{multline}
Observe that $d(t)=t/\sqrt{2}$ for $\gamma^-_x=0$, as expected. In the long time limit ($t/l\gg 1$), $ d(t)\simeq \tilde{m}(\gamma^- l)t$, with 
\begin{equation}
    \tilde{m}(\gamma^- l)=\sqrt{\frac{2}{\pi}}\bigg[\int_{0}^{1}dy\hspace{0.1cm}\frac{y^2}{\sqrt{1-y^2}}e^{-\gamma^- l/2y}\bigg]^{1/2}.
\end{equation}
In the quasi-particle picture, the transport is still ballistic but the average displacement cannot be linear in time anymore, since the dissipator may destroy the excitation in the interval $(-l/2,l/2)$. As expected, there is a crossover at $t=l/2$, coinciding with the distance covered by the particle with maximum velocity after the quench. In the long time limit, the average displacement is linear again. As before, we distinguish two asymptotic cases. For $\gamma^- l\ll 1$, $\exp{-\gamma^- l/2y}\simeq 1$ and $\tilde m\simeq 1/\sqrt{2}$; for $\gamma^- l\gg 1$, $\exp{-\gamma^- l/2y}\simeq 0$ and $\tilde m\simeq 0$.

\section{Wigner dynamics with dephasing}
\label{sec:dephasing}

\begin{figure}[t!]
\begin{center}
\includegraphics[width=\columnwidth]{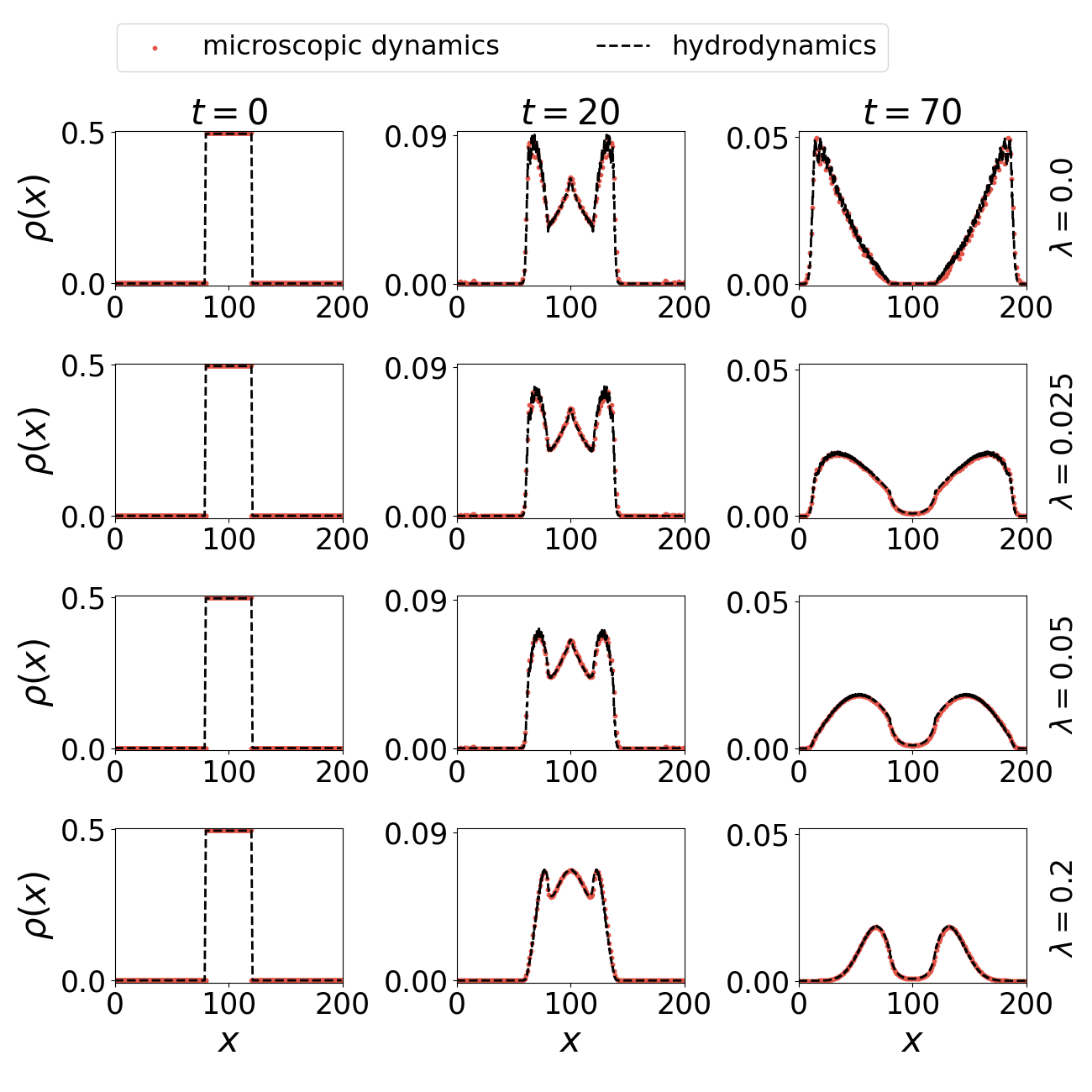}
     \caption{\label{density_plots_dep_and_loss}\textbf{Density time evolution - dephasing and loss processes:} 200-site chain of hopping fermions under homogeneous monitoring and local loss processes \eqref{gamma-}, $x_0=100$, $l=40$, $\gamma^-=0.1$. We prepare the system in the Wigner function \eqref{wigner_dep_loss} at time $t=0$. Here we show the density time evolution for different values of the parameter $\lambda$. The red spots and the black dashed lines refer to the numerical solution of the matrix differential equation \eqref{corr_mat_ev} and the hydrodynamic equation \eqref{Hydro_eq_4}, respectively.}
\end{center}
\end{figure}
\begin{figure}[t!]
\begin{center}
\includegraphics[width=\columnwidth]{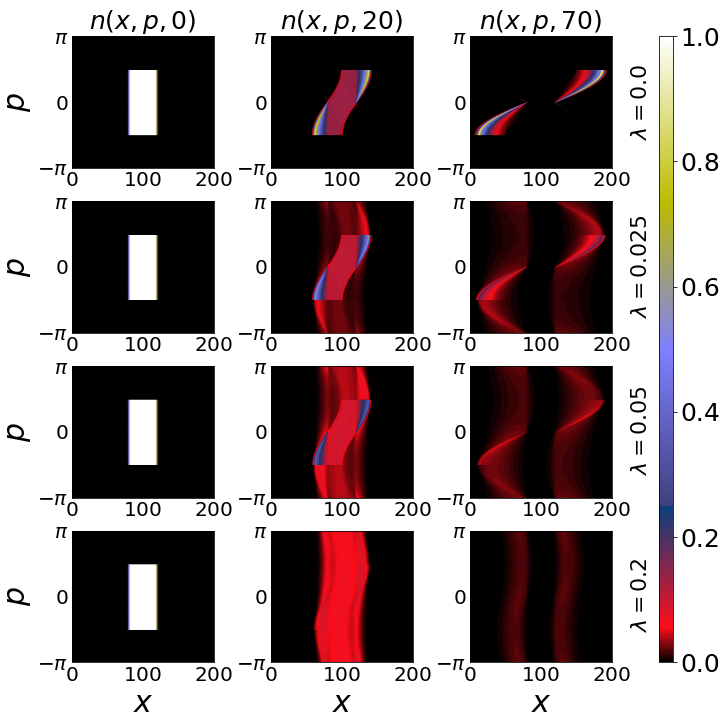}
\caption{\label{Wigner_plots_dep_and_loss}\textbf{Wigner function time evolution - dephasing and loss processes:} 200-site chain of hopping fermions under homogeneous monitoring and local loss processes \eqref{gamma-}, $x_0=100$, $l=40$, $\gamma^-=0.1$. We prepare the system in the Wigner function \eqref{wigner_dep_loss} at time $t=0$. In this figure, we show the full Wigner function time evolution for different values of the parameter $\lambda$. }
\end{center}
\end{figure}

In this section we include a constant dephasing in the Wigner function dynamics, characterizing continuous measurement processes, where the system undergoes infinitely weak and frequent interactions with ancillas \cite{breuer2002theory}.

Even if the pure dephasing dynamics has been abundantly explored, it is much less explicit how the combination of dephasing and gain/loss processes affects the Wigner evolution.

\subsection{Homogeneous dephasing and losses in real space}
For instance, we can imagine a one-dimensional system of hopping fermions (group velocity $v_p=\sin(p)$), under homogeneous dephasing and real-space losses only ($\lambda_x=\lambda$, $\gamma_x^+=0$, $\omega^+_p=\omega^-_p=\zeta_p=0$). In particular, we assume a single particle loss frequency $\gamma_x^-$ given by Eq.~\eqref{gamma-}, with $x_0=N/2$. Under these hypothesis, the Wigner function satisfies
\begin{equation}\label{Hydro_eq_4}
    \partial_t n(x,p,t)=-v_p\partial_x n(x,p,t)-(\lambda+\gamma^-_x)n(x,p,t)+\lambda\rho(x,t).
\end{equation}
We prepare the system in the Wigner function 
\begin{equation}\label{wigner_dep_loss}
    n(x,p,0)=\begin{cases}
    1\qquad \abs{x-N/2}<l/2 \land \abs{p}<\pi/2 \\
    0\qquad \abs{x-N/2}\geq l/2\vee\abs{p}\geq\pi/2
    \end{cases}
\end{equation}
and we study its time evolution. 

In Figs. \ref{density_plots_dep_and_loss}, \ref{Wigner_plots_dep_and_loss}, we show the density and the full Wigner function dynamics for several values of $\lambda$. An animation of the same process can be found in the supplemental material~\cite{SupMat}. In the large $\lambda$ limit, the Zeno regime freezes the dynamics and the system exponentially converges to the vacuum state. As the crossover time $\lambda^{-1}$ between the ballistic and the diffusive regime increases, more and more particles may escape from the chain region with $\gamma^-_x\neq 0$. In the limit $\lambda\to0$, the motion is purely ballistic and the number of destroyed particles is minimum. 
\begin{figure}[t]
\begin{center}
\includegraphics[width=\columnwidth]{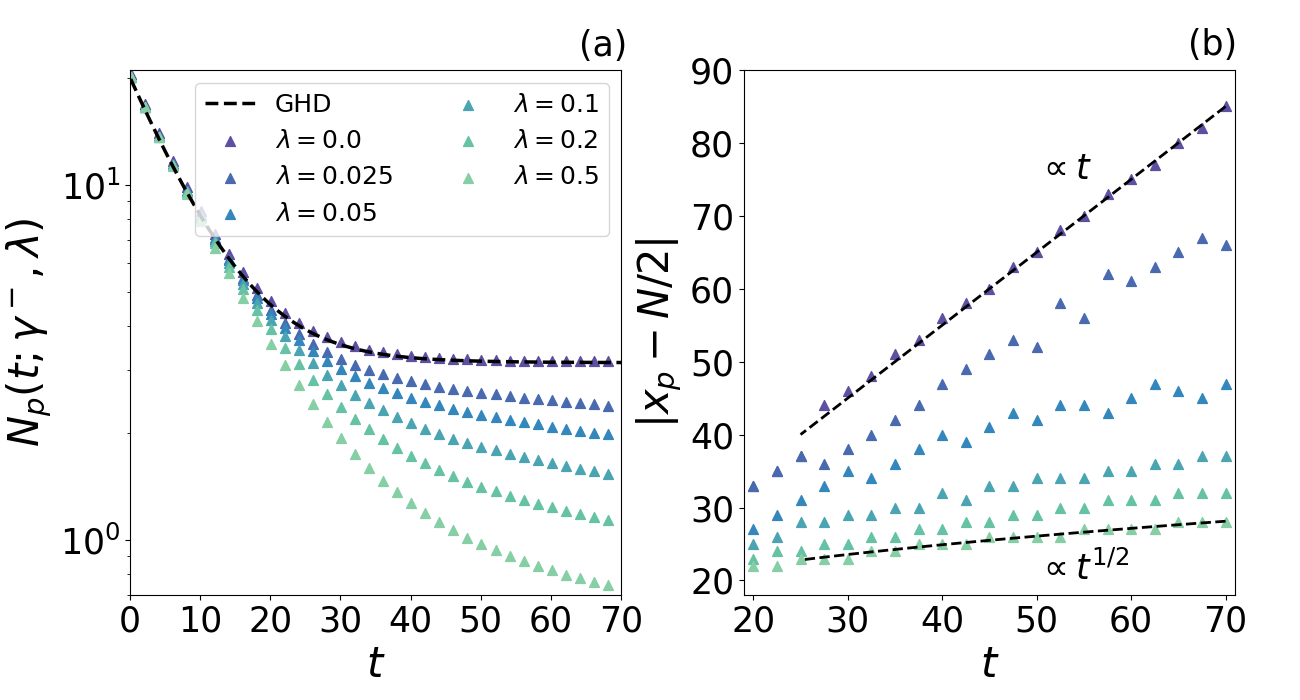}
     \caption{\label{peaks}\textbf{Total number of particles and peaks motion - dephasing and loss processes:} 200-site chain of hopping fermions under homogeneous monitoring and local loss processes \eqref{gamma-}, $x_0=100$, $l=40$, $\gamma^-=0.1$. We prepare the system in the Wigner function \eqref{wigner_dep_loss} at time $t=0$. (a) Total number of particles as function of time for different values of the parameter $\lambda$. The spots represent the data collected by solving Eq. \eqref{corr_mat_ev}. The black dashed line refer to the hydrodynamic prediction \eqref{numb_exact2}. (b) Displacement of the density peaks as function of time. We extrapolate the linear growth and the square root behavior.}
\end{center}
\end{figure}

\begin{figure}[t]
\begin{center}
\includegraphics[width=\columnwidth]{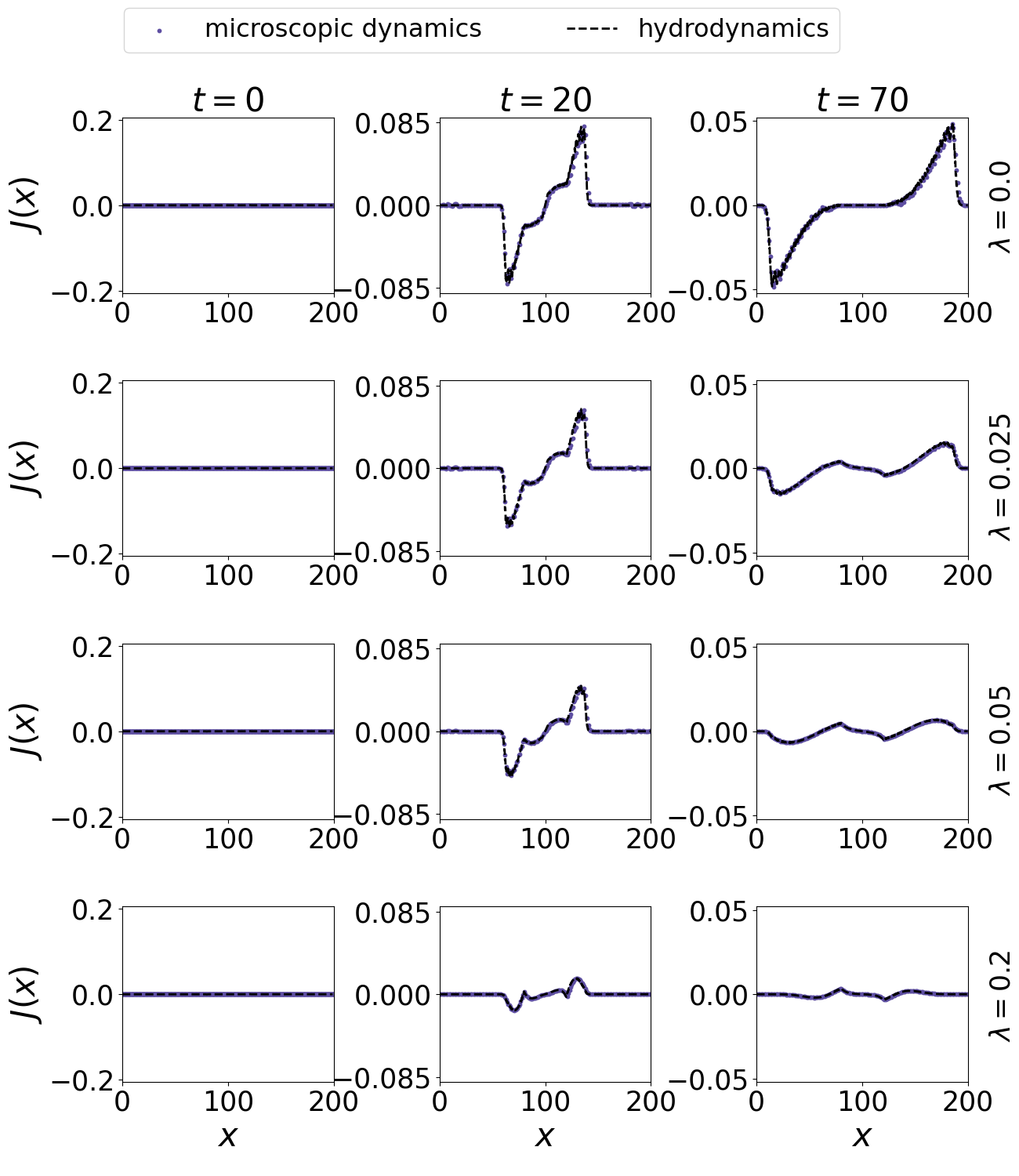}
\caption{\label{current_plots_dep_and_loss}\textbf{Current time evolution - dephasing and loss processes:} 200-site chain of hopping fermions under homogeneous monitoring and local loss processes \eqref{gamma-}, $x_0=100$, $l=40$, $\gamma^-=0.1$. We prepare the system in the Wigner function \eqref{wigner_dep_loss} at time $t=0$. Here we show the current time evolution for different values of the parameter $\lambda$. The blue spots represent the particle current from the numerical solution of the matrix differential equation \eqref{corr_mat_ev}. The black dashed lines refer to the hydrodynamic approach.}
\end{center}
\end{figure}

In Fig.~\ref{current_plots_dep_and_loss} we also show the current  $J(x,t)=\int_{-\pi}^{\pi}\frac{dp}{2\pi}\hspace{0.1cm}v_p n(x,p,t)$ at fixed time $t$ and measurement rate $\lambda$. It satisfies
\begin{equation}\label{cont_no}
    \partial_t \rho(x,t)+\partial_x J(x,t)+\gamma_x^-\rho(x,t)=0.
\end{equation}
As expected, the loss processes break the continuity equation. Integrating Eq. \eqref{cont_no} with the boundary condition $J(N,t)\xrightarrow{N\to\infty} 0$, we find 
\begin{equation}
    J(x,t)=\partial_t N_p([x,N],t)+\int_{x}^\infty dy\hspace{0.1cm} \gamma^-_{y}\rho(y,t),
\end{equation}
where $N_p([x,N],t)$ is the particle number in $[x,N]$ at time $t$. 

In Fig. \ref{peaks} (a), we plot the total number of particles $N_p(t;\gamma^-,\lambda)$ for different values of the parameter $\lambda$. $N_p(t;\gamma^-,\lambda)$ is a monotonically decreasing function of $\lambda$ for fixed values of $(t,\gamma^-)$. In the absence of dephasing ($\lambda=0$), 
\begin{multline}\label{numb_exact2}
    N_p(t;\gamma^-,0)=\Theta(l-t)\bigg[N_p^0e^{-\gamma^-t}-\frac{t}{\pi}e^{-\gamma^-t}+\frac{1}{\pi\gamma^-}(1-e^{-\gamma^-t})\bigg]\\
    +\Theta(t-l)\frac{1}{\pi\gamma^-}\bigg[1+l\gamma^-\arcsin\bigg(\frac{l}{t}\bigg)e^{-\gamma^-t}-e^{-\gamma^-t}(1+\gamma^- t)\bigg(1-\sqrt{1-\bigg(\frac{l}{t}\bigg)^2}\bigg)\\
    -\int_{\arcsin(l/t)}^{\pi/2}dp\hspace{0.1cm}\sin(p)e^{-\frac{\gamma^-l}{\sin(p)}}\bigg],
\end{multline}
where $N_p^0=l/2$ is the initial number of particles, according to the half filling condition \eqref{wigner_dep_loss}. If $t\to\infty$, the number of residual particles goes to $m(\gamma^-l)/\gamma^-$, approaching to $1/\pi\gamma^-$ for $\gamma^-l\ll 1$. In Fig. \ref{peaks} (b), we show $\abs{x_p-N/2}$ as function of time, where $x_p$ corresponds to the peak positions of the wave front of the density. As for the average displacement, we numerically find the initial linear regime and the $t^{1/2}$ behavior for $\lambda t\gg 1$. Observe that the position of the peaks (colored spots) in Fig. \ref{peaks} (b) is obtained by the microscopic dynamics. The higher order terms, which have been neglected after truncating Eq. \eqref{Hydro_eq}, are the reason behind the not complete monotonicity of the function $\abs{x_p-N/2}$. Indeed, this behavior cannot be captured by Eq. \eqref{Hydro_eq_4}. The purpose of Fig. \ref{peaks} (b) was to show a general tendency of the peaks to slow down for increasing dephasing constants, approaching the diffusive regime \eqref{diffusione} in the long time limit.
\subsection{Homogeneous dephasing and losses in momentum space}
\begin{figure}[t]
\begin{center}
\includegraphics[width=\columnwidth]{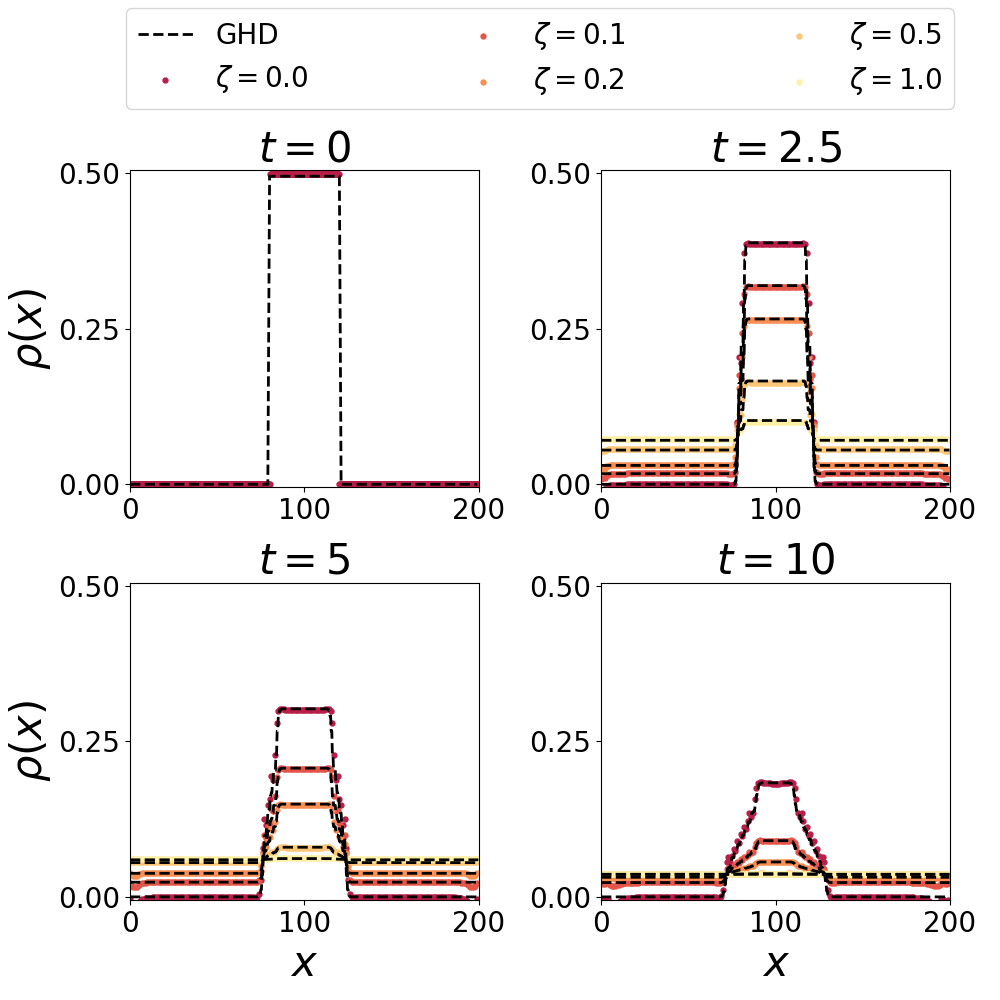}     \caption{\label{dens_four_plots_dep_and_loss}\textbf{Density time evolution - dephasing and loss processes:} 200-site chain of hopping fermions under homogeneous monitoring and local loss processes \eqref{jump_last}, $l=40$, $\omega^-=0.1$. We prepare the system in the Wigner function \eqref{wigner_dep_loss} at time $t=0$. Here we show the density time evolution for different values of the parameter $\zeta$. The colored spots and the black dashed lines refer to the numerical solution of the matrix differential equation \eqref{corr_mat_ev} and the hydrodynamic approach \eqref{Hydro_eq_5}, respectively.}
\end{center}
\end{figure}
In Fig. \ref{dens_four_plots_dep_and_loss}, we consider a chain of hopping fermions (group velocity $v_p=\sin(p)$), under homogeneous dephasing and momentum-space losses only ($\zeta_p=\zeta$, $\omega_p^+=0$, $\gamma^-_x=\gamma^+_x=\lambda_x=0$). The particle exchange is modulated by the jump frequency
\begin{equation}\label{jump_last}
\omega^-_p=\omega^-\bigg(\Theta\bigg(\frac{\pi}{2}-p\bigg)-\Theta\bigg(-\frac{\pi}{2}-p\bigg)\bigg),
\end{equation}
which selects the channels for the particle emission. 
Furthermore, the system is coupled to a monitoring apparatus which continuously and homogeneously measures the particle momentum. In this scenario, the Cauchy problem we want to solve is
\begin{equation}\label{Hydro_eq_5}
    \partial_t n(x,p,t)=-v_p\partial_x n(x,p,t)-(\zeta+\omega^-_p)n(x,p,t)+\zeta\tilde\rho(p,t),
\end{equation}
with the initial Wigner function \eqref{wigner_dep_loss}. In Fig. \ref{dens_four_plots_dep_and_loss}, we show the particle density evolution for different values of the monitoring rate $\zeta$. As the dynamics preserves the momentum conservation, the system exponentially converges to the vacuum state for any measurement rate. However, the larger parameter $\zeta$, the greater the number of delocalized particles per unit of time. 

\section{Discussion and conclusion}
\label{sec:disc}

In this work, we derived the hydrodynamics of quantum gas of  non-interacting particles coupled to external environments, including both linear and quadratic Lindblad operators. In particular, we found the partial differential equation for the Wigner function evolution under inhomogeneous jump rates. We described the dynamics in terms of classical quasi-particle motion: the gain and loss processes make the quasi-particles time of flight finite while the homogeneous dephasing affects the transport features, with a crossover from ballistic to diffusive regime. 
Through several examples, we showed how the Wigner dynamics perfectly captures all features of the full Lyapunov equation. 
Moreover, in several particular cases we have been able to provide analytical solutions, giving unique insights into the dynamics. Those results are clearly significant for transport phenomena in open systems, i.e. boundary driven quantum chains \cite{landi2021waiting}, offering new interesting perspectives to the understanding of the full counting statistics \cite{esposito2007fluctuation,esposito2009nonequilibrium,brandes2008waiting} and the waiting time distributions \cite{landi2021waiting}.
Finally, we studied the combined effects of constant dephasing and particle loss process. In this context, a natural working direction for the future would be to analyze the effects of a inhomogeneous monitoring rate, as for real experimental layouts, where more interesting transport features may emerge. 

\section*{Acknowledgments}
The authors acknowledge the financial support of the S\~ao Paulo Funding Agency FAPESP (Grant No.~2019/14072-0.), the Brazilian funding agency CNPq (Grant No. INCT-IQ 246569/2014-0) and the French ANR funding UNIOPEN (Grant No. ANR-22-CE30-0004-01). 

\bibliography{biblio}



\end{document}